# Impact of Generative AI (Large Language Models) on the PRA model construction and maintenance, observations.

**Valentin Rychkov**[*], **Claudia Picoco, Emilie Caleca**

[a]EDF R&D, Palaiseau 91120, France

**Abstract:**
The rapid development of Large Language Models (LLMs) and Generative Pre-Trained Transformers (GPTs) in the field of Generative Artificial Intelligence (AI) can significantly impact task automation in the modern economy. We anticipate that the PRA field will inevitably be affected by this technology[1]. Thus, the main goal of this paper is to engage the risk assessment community into a discussion of benefits and drawbacks of this technology for PRA. We make a preliminary analysis of possible application of LLM in Probabilistic Risk Assessment (PRA) modeling context referring to the ongoing experience in software engineering field. We explore potential application scenarios and the necessary conditions for controlled LLM usage in PRA modeling (whether static or dynamic). Additionally, we consider the potential impact of this technology on PRA modeling tools.

## 1. INTRODUCTION

Generative Artificial Intelligence (AI) and specifically rapid development of Large Language Models (LLM) and Generative Pre-Trained Transformers (GPT) may have a large impact on the automation of different tasks in modern economy. Despite a relatively young age of GPT technology [1], there already exist estimates of the impact of GPT services on the job quantity and quality (e.g., [2]).

In this paper, we explore potential implications of these technologies in the field of Probabilistic Risk Assessment (PRA). One can, for example, imagine using a specially trained generative AI model to query risks associated with a specific installation [3]. The goal would be to directly obtain risk insights from the AI. While a prototype of such a service may emerge soon, it's important to recognize that any AI application within a regulatory framework requires substantial justification to support any decision-making.

We try to imagine less disruptive scenario of generative AI application for PRA. Currently most of the risk assessment is based on modeling approaches. Therefore, in the following sections, we analyze the applicability of Generative AI in the PRA modelling process focusing on the possible applications during a PRA model development or maintenance.

Most of the risk assessment modeling for nuclear power plants is done using Boolean framework based on Event Trees (ET) and Fault Trees (FT). Dynamic risk assessment methodologies [4] present a promising complement to static Boolean framework. For example, an application of a dynamic risk assessment methodology in a regulatory framework has been recently reported [5,6].

From an analyst perspective ET/FT as well as modeling artefacts of different dynamic methodologies are graphical objects. Existing generative AI that can produce visual content [7] is not adapted for our needs. We want to treat both Boolean and dynamic models primarily as structured textual objects obeying certain syntax rules that can also be visualized. Once we treat a risk assessment model as a structured text, we can use a GPT to generate the risk assessment model (or its parts).

The most prominent use of GPT in a similar context is related to so-called programming copilots GitHub Copilot, CodeWhisperer, BingAI Chat (GPT 4.0) (see for detailed review [8]). Recent communication by Github paints a bright future for this type of technology [9]. On the contrary, academic research suggests that a close attention must be given to the quality of the generated code [10-12].

The paper faces the challenge of reconciling opposing factors:

---

[1] We could have entitled the paper "The Large Language Models for PRA: not if but when", but the copyright for it belongs to Nathan SIU for his technical opinion paper on Dynamic PRA.

- It aims to provide well-founded observations about the technology that authors are not experts in, despite conflicting claims from vendors and the research community.
- The paper discusses the integration of rapidly advancing technology - where developments from just a year ago may already be outdated - into the traditionally cautious and slowly evolving nuclear field, particularly in probabilistic risk and safety assessment, where small changes can take years to accept.

The paper is organized as follows: in Section 2 we first briefly discuss current application scenarios of generative AI for the software engineering, in Section 3 and Section 4 we discuss in detail if these scenarios are relevant for the PRA modeling in static and dynamic context, respectively. In Section 5 we discuss the challenges related to verification of static and dynamic models. In Section 6, we describe possible implications for PRA tools. In section 7 we provide a list of other important questions that are not covered in the paper. Finally, some conclusions are drawn in Section 8.

## 2. PROS AND CONS OF GENERATIVE AI IN SOFTWARE ENGINEERING.

The state of practice of the generative AI in software engineering is well summarized by Baskhad Idrisov and Tim Schlippe in [8]:

*"AI-powered chatbots are not only capable of conducting human-like conversations but can also generate program code […] Simultaneously, Integrated Development Environments (IDEs), unit tests, and benchmarking tools have simplified coding, making it more accessible to both seasoned developers and Beginners."*

Recent communication [9] by Github provides the following claims:
- *85% of developers felt more confident in their code quality when authoring code with GitHub Copilot and GitHub Copilot Chat.*
- *Code reviews were more actionable and completed 15% faster with GitHub Copilot Chat.*
- *88% of developers reported maintaining flow state with GitHub Copilot Chat because they felt more focused, less frustrated, and enjoyed coding more, too.*

In the communication [13], Github claimed that the developers using GitHub Copilot code 55% faster.

The academic research provides less bright picture. While AI tools promise to facilitate the coding process and lower the barriers to entry for the beginners, it is important to remember that these tools still have their limits [8], they may not always produce correct, efficient, or maintainable program codes. The recent empirical study [12] finds that that 52% of ChatGPT answers to 517 programming questions at Stack Overflow contain incorrect information and 77% were verbose. For more discussion and examples refer to [10-12].

In summary, the generative AI tools could provide a substantial productivity gain for coding, but a particular attention must be given to different quality metrics: correctness, performance, and maintainability. In the following sections we discuss if these expectations can be transposed to PRA modeling domain.

## 3. APPLICABILITY OF GENERATIVE AI FOR EVENT TREE AND FAULT TREE MODELING.

Boolean methods (Event Trees and Fault Trees) have been widely used for the risk assessment since 1962 [14]. This paper is not intended to provide an exhaustive overview of the evolution of the ET/FT modeling practices. Nevertheless, we highlight the efforts that have been made to improve the efficiency of the process.

In simple terms, a Fault Tree can be seen as a very specific representation of a system that aims to explain how a system function can fail given the failures (or more generally specific states) of its subsystems. A system can fulfil different functions; thus, it can be mapped on different fault trees.

In practice, to construct a fault tree for a given system function a reliability engineer has three options:
1. construct it manually (graphically or in textual mode using e.g. OpenPSA [15]) using logical gates and basic events.
2. write a representation of the system in a dedicated reliability modeling language [16,17] that can be converted into a fault tree.
3. use a Piping and Instrumentation Diagram PID diagram that is later mapped to the fault tree.

The three options mentioned above assume that the created fault tree is ultimately represented in a format compatible with a fault tree quantification software. Options 1 and 2 imply a manual transformation of the available information about the system into a fault tree. Option 3 presents more automated solution. Historically at EDF the KB3 or RISKSPECTRUM ModelBuilder software [18] have been developed to construct "Piping and Instrumentation (PID) like" diagrams that are automatically converted into the fault trees. Recently a more automated solution that leverage existing PID diagrams used for the system design have been proposed [19,20].

A similar solution exists for the Event Trees generation from functional block diagrams [21] and the link with Emergency Operating Procedures (EOPs) is being explored [22]. These efforts aim to improve the PRA modeling efficiency, reproducibility, and maintainability.

Today the PRA community may be satisfied with obtained results, but nothing prevents us from looking for other better or/and more cost-effective options to achieve modeling tasks That is why in the following sections we imagine different application scenarios of generative AI for PRA modeling.

1. **A generative AI use case for fault tree modeling**

Below we provide an example that illustrates the result of a possible interaction[2] of a PSA engineer with a FaultTree generation copilot (Microsoft Copilot have been used for this example).

**Query 1:**
*Please generate a fault tree in OpenPSA format for the system that*
- *has three redundant components.*
- *two out of three components are necessary to fulfill the system function.*

**Answer 1**:

```xml
<fault_tree>
  <!--Basic events (components) -->
  <basic_event name="Component_A" />
  <basic_event name="Component_B" />
  <basic_event name="Component_C" />
  <!--Gates -->
  <gate type="AND" name="AND1">
    <input>Component_A</input>
    <input>Component_B</input>
  </gate>
  <gate type="AND" name="AND2">
    <input>Component_A</input>
    <input>Component_C</input>
  </gate>
  <gate type="OR" name="OR1">
    <input>AND1</input>
    <input>AND2</input>
  </gate>
</fault_tree>
```

Code 1. GPT generated FT.

First, we note that AI chatbot correctly understood the query and provided a plausible answer. In addition to a generated fault tree there have been provided a small text (omitted in the paper) that supposed to represent a rational behind the answer.

Second, we note that the provided answer is not correct. One combination of failures is missing. The following indication was provided to the chatbot:

**Query 2:**
*One combination of failures is missing in your answer, could you please correct it.*

---

[2] Authors are aware of different techniques of prompt engineering, that can improve the supposed precision of generative AI answer, but it is out of scope of this paper, see Section 6.

**Answer 2:**

```xml
<fault_tree>
  <!--Basic events (components) -->
  <basic_event name="Component_A" />
  <basic_event name="Component_B" />
  <basic_event name="Component_C" />
  <!--Gates -->
  <gate type="AND" name="AND1">
    <input>Component_A</input>
    <input>Component_B</input>
  </gate>
  <gate type="AND" name="AND2">
    <input>Component_A</input>
    <input>Component_C</input>
  </gate>
  <gate type="AND" name="AND3">
    <input>Component_B</input>
    <input>Component_C</input>
  </gate>
  <gate type="OR" name="OR1">
    <input>AND1</input>
    <input>AND2</input>
    <input>AND3</input>
  </gate>
</fault_tree>
```

Code 2. GPT generated FT, corrected.

**Additional observations:**

We performed a few additional experiments reformulating the first query and the results were not consistent[3]. One of the experiments led to a fault tree that had four different basic events representing four different components. At a different attempt we got a fault tree that represented a system that could function with 1 component out of 3 (triple redundancy):

```xml
<FT>
  <Event name="SystemMissionFailure" type="TopEvent">
    <Gate type="AND">
      <Event name="Component1Failure" type="BasicEvent" />
      <Event name="Component2Failure" type="BasicEvent" />
      <Event name="Component3Failure" type="BasicEvent" />
    </Gate>
  </Event>
</FT>
```

Code 3. GPT generated FT for a redundancy 3 system

With these examples we do not aim to illustrate an obvious fact that a generative AI can provide an incorrect output, but rather draw an attention to the importance of the verification of the obtained results.

**Observation 1**: to allow the generative AI use in the PRA modeling context one needs to develop automatic verification methods and tools for PRA artefacts: such as Fault Trees and Event Trees.

## 2. A generative AI use case for a fault tree documentation

Generative AI have been also used in software engineering to provide an explanation of a given part of the programming code in the natural language [23].

---

[3] After the initial publication, authors have been notified that other (the most recent) GPT technologies perform much better with the presented tests. Correct answers were obtained for all presented examples from the first attempt.

If we transpose that into PRA modeling world, we can imagine different use cases. Below we provide two obvious application areas.
1. Documentation generation: lack of documentation is known to be a generic issue for many domains. Specifically in the software engineering, as in this paper we try to provide the close analogies with the domain, the lack of documentation leads to, so called, technological debt and maintainability problems [24]. In that context, the PRA models are not that much different from programming code. Despite sometimes regulatory requirements [25] on the documentation side, one rarely encounters a situation when there is too much model documentation. We can imagine a generative AI as a support to complete the gaps in the existing documentation based on the existing model. An example of such a use case will be provided below.
2. A reverse PID generation from a fault tree: one can think of another possible use case of generative AI in the context of Fault Tree interpretation, a PID generation from the Fault Trees. This functionality can be useful to generate additional documentation for legacy PRA models. This functionality can also be useful in for the verification of Fault Tree results. We mention that an attempt of such a conversion have been already made without the use of generative AI [26]. Thus, a cross verification of AI generated PID with rule-based algorithms can provide a solid verification base.

Below we provide an example that illustrates the result of an interaction of an engineer with "a FautTree documentation copilot" (Microsoft Copilot have been used for this example). We asked the GPT service to provide an explanation of a given fault tree in an unstructured language.

**Query 1:**
*Provide an explanation of system functioning described by a FT [see Code 3].*
**Answer 1:**
1. ***SystemMissionFailure (Top Event):*** *This represents the overall system failure, which occurs if any of the following three components fail simultaneously:*
    - ***Component1Failure***
    - ***Component2Failure***
    - ***Component3Failure***
2. ***Gate Type: AND (Conjunction):*** *The AND gate indicates that all three components must fail for the top event (SystemMissionFailure) to occur. In other words, it's a logical AND operation.*
3. ***Basic Events:***
    - ***Component1Failure:*** *Represents the failure of the first component.*
    - ***Component2Failure:*** *Represents the failure of the second component.*
    - ***Component3Failure:*** *Represents the failure of the third component.*

*To summarize, the fault tree describes a system where the mission fails if all three components (Component1, Component2, and Component3) fail simultaneously.*

We note that the answer is satisfactory and correct. A similar exercise was performed with the Code 2 example with the same level of satisfaction in the obtained answers.

**Query 1a:**
*Provide an explanation of system functioning described by a FT [see Code 2].*

**Answer 1a:**
- *The system will function if either:*
    - ***Component_A*** *and* ***Component_B*** *are both operational, or*
    - ***Component_A*** *and* ***Component_C*** *are both operational, or*
    - ***Component_B*** *and* ***Component_C*** *are both operational.*
- *If all three paths fail (i.e., both* ***Component_A*** *and* ***Component_B***, ***Component_A*** *and* ***Component_C***, *and* ***Component_B*** *and* ***Component_C*** *fail), the system will fail.*

**Query 2:**
*What can you say about the level of redundancy of system described by a FT [see Code 3]?*

**Answer 2:**
*Let's analyze the level of redundancy in the system described by the fault tree.*
1) ***Components:*** *The system consists of three components: Component1, Component2, and Component3.*
2) ***Redundancy:***

i) The AND gate in the fault tree indicates that all three components must fail simultaneously for the top event (SystemMissionFailure) to occur.
ii) This configuration does not provide any inherent redundancy because there is no backup or alternative path. If any one of the components fails, the entire system mission fails.

We note that the provided answer is not satisfactory as 2.i and 2.ii contradict to each other. The same question about the redundancy was asked about the system characterized by the fault tree described in Code 2. Satisfactory explanation was obtained.

**Query 2a:**
*What can you say about the level of redundancy of system described by a FT [see Code 2]?*

**Answer 2a**
*Let's assess the redundancy in this system:*
- ***Component_A***:
  - *Appears in both **AND1** and **AND2** paths.*
  - *Provides redundancy because it is part of two separate paths.*
- ***Component_B***:
  - *Appears in both **AND1** and **AND3** paths.*
  - *Provides redundancy because it is part of two separate paths.*
- ***Component_C***:
  - *Appears in both **AND2** and **AND3** paths.*
  - *Provides redundancy because it is part of two separate paths.*

*Overall, this system has a **redundancy level of 2** because each component is part of two independent paths.*

Again, we don't intend to point out errors and inconsistencies in an AI generated documentation. We rather want to start the discussion about possible implication of generative AI in documentation of existing and future PRA models as well as necessary technological and organizational conditions that will allow to leverage obvious advantages of this technology. Indeed, to foresee such an application of Generative AI for documentation, one should define what is the content of model documentation and train the AI accordingly. For example, if the documentation consists of tracking modeling assumption, the Generative AI should be able either to "detect" assumptions or the lack of them in the documentation and to point them to the analyst.

**Observation 2**: We think that already at the current state, the generative AI can be used as an assistant to enrich existing/developing PRA model documentation, under the condition that: (i) one defines what the documentation consist of and (ii) the AI generated documentation is reviewed by engineers.

### 4. A generative AI use case for a faut tree review.

In §3.1 and §3.2 we discussed how a generative AI can be used to generate a fault tree starting from a natural language and generate a natural language documentation for a given fault tree. In this section we provide an example that combines both use cases, namely we use a generative AI to do a review of a fault tree to detect if the natural language description corresponds to its faut tree representation.

**Query 1:**
*Please find a possible error in the fault tree in the OpenPSA format that corresponds to the following system description.*
[SYSTEM DESCRIPTION]
*- has three redundant components.*
*- two out of three components are necessary to    ulfil the system function.*
[/SYSTEM DESCRIPTION]
[OPENPSA FAULT TREE]
*See Code 1*
[/OPENPSA FAULT TREE]

**Answer 1:**
*The fault tree you provided does correspond to the system description. Here's why:*
- *The system has three redundant components: **Component_A**, **Component_B**, and **Component_C**.*
- *The system function is fulfilled if at least two out of these three components are functioning. This is represented by the AND gates (AND1 and AND2) and the OR gate (OR1) in the fault tree.*
- *Specifically, the system function is fulfilled if either of the following conditions is met:*
  - ***Component_A** and **Component_B** are functioning (represented by AND1)*

- *Component_A and Component_C are functioning (represented by AND2)*
- *This configuration ensures that the system can still function even if one component fails, which aligns with the system description you provided. Therefore, the fault tree correctly represents the system. There doesn't seem to be any error in the fault tree based on the given system description.*

We note that the provided answer is not satisfactory, as we already discussed in 3.1. In the following query we slightly reformulate (adding one word) the system description to test the sensitivity of the AI answer

**Query 2:**
*Please find a possible error in the fault tree in the OpenPSA format that corresponds to the following system description*
[SYSTEM DESCRIPTION]
*- has three redundant components.*
*- any two out of three components are necessary to    ulfil the system function.*
[/SYSTEM DESCRIPTION]
[OPENPSA FAULT TREE]
*See Code 1*
[/OPENPSA FAULT TREE]
**Answer 2:**
*The fault tree you provided does not fully correspond to the system description. Here's why:*
- *The system has three redundant components: **Component_A**, **Component_B**, and **Component_C**. The system function is fulfilled if any two out of these three components are functioning. However, the fault tree only considers two specific combinations of components: (**Component_A** and **Component_B**) or (**Component_A** and **Component_C**).*
- *The fault tree does not account for the possibility of **Component_B** and **Component_C** functioning while **Component_A** fails. This is a possible scenario according to your system description, but it is not represented in the fault tree.*

*To correct this, you would need to add another AND gate for **Component_B** and **Component_C**, and include this in the final OR gate. Here's how it could look [see Code 2]*

The copilot provided the correct answer once we made a tiny adjustment to the system description. If the sensitivity of the model to such a tiny reformulation of the query cannot be excluded, it can hardly, if ever, be coped with a specific training of the user (see open questions in the **Section 6**).

**Observation 3:** A generative AI can be used as a support to review of a PRA model. Using specifically trained AI one can probably achieve high error/inconsistency detection rate in a PRA model providing valuable support for reviewer team. The following questions remains to be answered: what is the minimum error detection rate to leverage the advantages of the technology? What is the admissible false positive error detection rate? How do we judge a model where all AI detected errors were fixed?

## 5. APPLICABILITY OF GENERATIVE AI FOR DYNAMIC RELIABILITY MODELING.

Compared to Boolean PRA where we used a general purpose trained LLM (Microsoft Copilot) to test their applicability, dynamic reliability modeling can already benefit from existing specific solutions.

There are a substantial number of dynamic methodologies of different degree of maturity that have been developed over last 30 years.

Recently, authors proposed to use software engineering techniques for Dynamic PRA [27,28]. More specifically, we adapted statechart visual formalism [29, 30] for reliability domain.

There exist several tools that support statechart modeling e.g., itemisCREATE, Rational Rhapsody by IBM, IAR visual state, stateFlow for MATLAB, stately.ai (see [31] for a detailed review), and there already exist/announced two dedicated AI powered copilots from itemis [32] and from stately [33] to assist statechart modeling.

In this paper we do not intend to provide a demonstration of the AI capabilities for state based (statechart) modeling as it is already on the market. Application cases discussed in 3.1, 3.2, and 3.3 are also relevant for dynamic PRA modeling. The demonstration of the discussed capabilities can be found, e.g., at the itemis web page.

Instead, we would like to point out important differences between static and dynamic models that may favor the deployment of AI powered solution for one than for another.

## 6. SYNTAX, SEMANTICS AND RIGOROUS VERIFICATION.

Picoco et al already discussed [34] the difference in the verification approaches for Boolean and Dynamic modeling. In essence, the main difference between Boolean and Dynamic models is that a Dynamic model is executable, while a Boolean model is not. A more formal description of the differences is that a dynamic model must have a formally defined syntax and semantics, while a Boolean model needs to have a defined syntax only[4].

As we discussed previously (**Observation 1** and **2**) it is very risky to trust an answer form a generative AI without means to verify its correctness. It is relatively easy to verify if a model that represents a Boolean formula is correct from a grammatical point of view. Authors are not aware of research about semantic verification of Boolean PRA models. From our point of view, the "semantic" verification of a Boolean PRA model is done through a peer review and expert judgment only [35]. An AI assisted review (**Observation 3**) can be seen as a complement to the expert review, nevertheless it does not represent a rigorous semantic verification.

On the other side, dynamic PRA models can be verified for grammar and semantics with different techniques [36] that allow to better judge the correctness of the generative AI output.

**Observation 4**: From our perspective, PRA community paid insufficient attention to automatic verification of PRA models. It is worth noticing, when it comes to verification, dynamic models may have (depending on the considered approach) an advantage over static models. Authors suggest that the community invest more energy to adopt existing software engineering verification methodologies to dynamic modeling as well as to develop necessary approaches to verify Boolean models to fully leverage advantages of generative AI.

## 6. TOOLS AND EXCHANGE FORMATS

The experiments that we were able to conduct with Fault Tree generation using Microsoft Copilot were only possible because the OpenPSA format is an open exchange format. It seems that the training set of the LLM behind the GPT service contained several OpenPSA related datapoints.

Given the current state of GPT technology we present two possible scenarios: the PRA software developers will either
   a) be led to adopt open exchange formats for the PRA models to share the costs related with the training/implementation/maintenance of the generative solutions.
   b) be able to conserve their proprietary formats customizing existing AI solutions[5].

## 7. OTHER IMPORTANT TOPICS NOT COVERED IN THE PAPER
Authors are aware of multiple other topics not covered in the paper, just to cite a few:
   - Data security: how do we ensure the sensitive data security when training/using a PRA oriented AI?
   - Economics: What is the cost to obtain a specifically trained generative AI or to adopt an existing one in an industrial environment? How much of the productivity gain it brings for different use cases?
   - Common causes: If different utilities use commonly trained AI powered solution is there the risk to overlook the same vulnerabilities?
   - Organization adjustments: how much of specific training is needed for people who are going to use AI based solutions for PRA?
   - Regulatory perspective and policy: Are the imagined use cases compatible with existing and coming policies?
   - Knowledge management: how the role of an expert changes in the presence of an AI based solution, should one be afraid of general knowledge degradation?

## 8. CONCLUSIONS

---

[4] Syntax is the concept that concerns itself only whether a "sentence" is valid for the grammar of the language. Semantics is about whether the "sentence" has a valid meaning.

[5] See for example https://en.wikipedia.org/wiki/Generative_pre-trained_transformer#Domain-specificity

In this paper we tried to imagine application scenarios of generative AI (LLM, GPT) in the context of PRA modeling. We formulate the following observations:

**Observation 1**: to allow the generative AI use in the PRA modeling context one needs to develop automatic verification methods and tools for PRA artefacts.

**Observation 2**: We think that already at the current state, the generative AI can be used as an assistant to enrich existing/developing PRA model documentation, under the condition that the AI generated documentation is reviewed by engineers.

**Observation 3:** A generative AI can be used as a support to review of a PRA model. Using specifically trained AI one can probably achieve high error/inconsistency detection rate in a PRA model providing valuable support for reviewer team. Some questions remain to be answered, for example what is the minimum error detection rate to leverage the advantages of the technology? What is the admissible false positive error detection rate? How do we judge a model where all AI detected errors were fixed?

**Observation 4**: From our perspective, PRA community paid insufficient attention to automatic verification of PRA models. It is worth noticing, when it comes to verification, dynamic models may have (depending on the considered approach) an advantage over static models. Authors suggest that the community invest more energy to adopt existing software engineering verification methodologies to dynamic modeling as well as to develop necessary approaches to verify Boolean models to fully leverage advantages of generative AI.

The paper concerns rapidly evolving domain. Imagined application scenarios can change with the growing practice of application of generative AI in other domains, thus the observations presented here may become obsolete soon.

**References**
[1] Vaswani, A et al. "Attention is All You Need". In Advances in Neural Information Processing Systems; MIT Press: Cambridge, MA, USA, (2017); p. 30.
[2] Brynjolfsson, E et al. *Generative AI at work*. No. w31161. National Bureau of Economic Research, 2023.
[3] Y. Hu, et al, "The Use and Misuse of Pre-Trained Generative Large Language Models in Reliability Engineering," *2024 Annual Reliability and Maintainability Symposium (RAMS)*, Albuquerque, NM, USA, 2024, pp. 1-7, doi: 10.1109/RAMS51492.2024.10457630.
[4] Tunc A, "A survey of dynamic methodologies for probabilistic safety assessment of nuclear power plants", *Annals of Nuclear Energy*, Volume 52, 2013, https://doi.org/10.1016/j.anucene.2012.08.001.
[5] Moulié F., et al. Analyses Probabilistes de Sûreté innovantes avec modélisations dynamiques. Congrès Lambda Mu 23 " Innovations et maîtrise des risques pour un avenir durable " - 23e Congrès de Maîtrise des Risques et de Sûreté de Fonctionnement, IMDR, Oct 2022, Paris Saclay, France. hal-03966660
[6] AVIS IRSN N° 2024-00049 « Établissement Orano Recyclage de La Hague - INB n 116 (UP3-A) et INB n 117 (UP2-800) Etudes probabilistes de sûreté en réponse aux engagements n 9 et n 17 pris dans le cadre des réunions du groupe permanent d'experts pour les laboratoires et les usines relatives au réexamen périodique de l'usine UP2-800 », https://www.irsn.fr/sites/default/files/2024-05/Avis-IRSN-2024-00049.pdf Accessed 05.2024
[7] https://openai.com/index/dall-e-2/ Accessed 05.2024
[8] Idrisov, B.; Schlippe, T. Program Code Generation with Generative AIs. Algorithms 2024, 17, 62. https://doi.org/10.3390/a17020062
[9] https://github.blog/2023-10-10-research-quantifying-github-copilots-impact-on-code-quality/ Accessed 05.2024
[10] Zhang, B.; Liang, P.; Zhou, X.; Ahmad, A.; Waseem, M. Practices and Challenges of Using GitHub Copilot: An Empirical Study. In Proceedings of the International Conferences on Software Engineering and Knowledge Engineering, San Francisco, CA, USA, (2023)
[11] Liu, Jiawei, et al. "Is your code generated by chatgpt really correct? rigorous evaluation of large language models for code generation." Advances in Neural Information Processing Systems 36 (2024)
[12] Samia Kabir, et al. "Is Stack Overflow Obsolete? An Empirical Study of the Characteristics of ChatGPT Answers to Stack Overflow Questions". In Proceedings of the CHI Conference on Human Factors in Computing Systems (CHI '24), May 11–1 (2024)
[13] https://github.blog/2022-09-07-research-quantifying-github-copilots-impact-on-developer-productivity-and-happiness/ Accessed 05.2024
[14] Ericson, C.. *"Fault Tree Analysis - A History"* . Proceedings of the 17th International Systems Safety Conference. (1999).

[15] M Hibti, et al. "*Overview of the open psa platform*" - Proceedings of International Joint Conference PSAM, 2012

[16] Prosvirnova, T., et al. "The AltaRica 3.0 project for model-based safety assessment." *IFAC proceedings volumes* 46.22 (2013): 127-132.

[17] Bouissou, M., Villatte, N., Bouhadana, H., & Bannelier, M. Knowledge modelling and reliability processing: presentation of the FIGARO language and associated tools (No. EDF--92-NB-00028). Electricite de France (EDF) (1991).

[18] Renault, I., Pilliere, M., Villatte, N., & Mouttapa, P. KB3: computer program for automatic generation of fault trees. Annual Reliability and Maintainability. Symposium. (1999) Proceedings (Cat. No.99CH36283). https://doi.org/10.1109/rams.1999.744149

[19] Roy, R., Houdebine J.C., and Hibti M. "Récupération automatique des données de conception des systèmes pour les études de fiabilité." Congrès Lambda Mu 23 «Innovations et maîtrise des risques pour un avenir durable»-23e Congrès de Maîtrise des Risques et de Sûreté de Fonctionnement, Institut pour la Maîtrise des Risques. (2022).

[20] Roy, R., Houdebine J.C., and Hibti M. «Illustration d'exemples de récupération automatique de données de conception pour les études de sûreté avec représentation compressée et/ou par attribut et/ou par multi configuration ». to appear in" Congrès Lambda Mu 24 «Innovations et maîtrise des risques pour un avenir durable»-24e Congrès de Maîtrise des Risques et de Sûreté de Fonctionnement, Institut pour la Maîtrise des Risques. (2024)

[21] Friedlhuber T., Hibti M., and Rauzy M.. Automated generation of event trees from event sequence functional block diagrams and optimisation issues. In Proceedings of PSAM Topical Conference in Tokyo, April (2013)

[22] Roy, R., Massoulier, C. « Mise en œuvre d'une méthode pour visualiser de manière synthétique les procédures de conduite incidentelle et accidentelle avec la vision des boucles pour les études probabilistes de sûreté » to appear in" Congrès Lambda Mu 24 «Innovations et maîtrise des risques pour un avenir durable»- 24e Congrès de Maîtrise des Risques et de Sûreté de Fonctionnement, IMDR. 2024

[23] MacNeil, Stephen, et al. "Generating diverse code explanations using the gpt-3 large language model." Proceedings of the 2022 ACM Conference on International Computing Education Research-Volume 2. 2022.

[24] Rios, Nicolli, et al. "Hearing the voice of software practitioners on causes, effects, and practices to deal with documentation debt." Requirements Engineering: Foundation for Software Quality: 26th International Working Conference, REFSQ 2020, Pisa, Italy, March 24–27, 2020, Proceedings 26. Springer International Publishing, 2020.

[25] Probabilistic Safety Analysis ONR Guide CM9 2020/262092, page 74 https://www.onr.org.uk/media/msjpkl10/ns-tast-gd-030.pdf

[26] Bos, Wouter, et al. "From Fault Trees to Piping and Instrumentation Diagrams." 33rd European Safety and Reliability Conference, ESREL 2023. Research Publishing, (2023.)

[27] Rychkov V., Picoco C., "Model based software engineering techniques for dynamic reliability assessment", ESREL2022, Dublin, (2022)

[28] Rychkov, V., et al. "Statecharts as a Dynamic Method for Risk Assessment.", Topical Meeting on Probabilistic Safety Assessment, Knoxville USA (2023)

[29] Harel, D. "Statecharts: A visual formalism for complex systems". *Science of computer programming*, *8*(3), 231-274. (1987)

[30] https://www.edx.org/learn/computer-programming/israelx-programming-for-everyone-an-introduction-to-visual-programming-languages  Accessed 05.2024

[31] Spanoudakis, Nikolaos I. "Engineering Multi-agent Systems with Statecharts: Theory and Practice." SN Computer Science 2.4 (2021): 317

[32] https://www.itemis.com/en/products/itemis-create/key-features/#ai-supported-modeling  Accessed 05.2024

[33] https://stately.ai/ Accessed 05.2024

[34] Picoco C., Rychkov V., Aldemir T., "A framework for verifying Dynamic Probabilistic Risk Assessment models", Reliability Engineering & System Safety, (2020).

[35] Finnicum D. Regulatory guide 1.200 pra peer reviews: a peer reviewer perspective. ASME. international conference on nuclear engineering, Volume 1: plant operations, maintenance, installations and life cycle; component reliability and materials issues; advanced applications of nuclear technology; codes, standards, licensing and regulatory issues. 2008. p. 921

[36] André, Étienne, et al. "Formalizing UML state machines for automated verification–A survey." *ACM Computing Surveys* 55.13s (2023): 1-47.